\documentclass[showpacs,showkeys,12pt,preprint,preprintnumbers,nofootinbib,
groupedaddress,epsfig,superscriptaddress,amsmath,amssymb]{revtex4}

\usepackage{graphicx}
\usepackage{dcolumn}
\usepackage{bm}
\usepackage{amsmath}
\usepackage{epsfig} 
\usepackage{here}

\def\beq{\begin{equation}}
\def\eeq{\end{equation}}
\def\bea{\begin{array}}
\def\eea{\end{array}}
\def\be{\begin{equation}}
\def\ee{\end{equation}}
\def\ba{\begin{eqnarray}}
\def\ea{\end{eqnarray}}

\def\to{\rightarrow}

\def\[{\left[}
\def\]{\right]}
\def\({\left(}
\def\){\right)}

\def\sm0{{\widetilde{m}_0}}

\def\U1em{{U(1)_{\rm em}}}
\def\to{\rightarrow}

\def\sq2{\sqrt{2}}

\def\ee{e^+e^-}

\def\End{\end{document}}

\newcommand{\gsim}{\mbox{ \raisebox{-1.0ex}{$\stackrel{\textstyle >}
{\textstyle \sim}$ }}}

\def\fsl#1{\setbox0=\hbox{$#1$}                 
   \dimen0=\wd0                                 
   \setbox1=\hbox{/} \dimen1=\wd1               
   \ifdim\dimen0>\dimen1                        
      \rlap{\hbox to \dimen0{\hfil/\hfil}}      
      #1                                        
   \else                                        
      \rlap{\hbox to \dimen1{\hfil$#1$\hfil}}   
      /                                         
   \fi}

\begin{document}

\title{Triviality and vacuum stability bounds in the three-loop \\ neutrino mass model }
\author{Mayumi Aoki}
\email{mayumi@hep.s.kanazawa-u.ac.jp}
\affiliation{Institute~for~Theoretical~Physics,~Kanazawa~University,~Kanazawa~920-1192, ~Japan}
\author{Shinya Kanemura}
\email{kanemu@sci.u-toyama.ac.jp}
\affiliation{Department~of Physics,~University~of~Toyama,~3190~Gofuku,~Toyama~930-8555,~Japan}
\author{Kei Yagyu}
\email{keiyagyu@jodo.sci.u-toyama.ac.jp}
\affiliation{Department~of Physics,~University~of~Toyama,~3190~Gofuku,~Toyama~930-8555,~Japan}
%
\preprint{KANAZAWA-11-01, UT-HET 035}
\pacs{\, } 

\begin{abstract}
We study theoretical constraints on the parameter space under 
the conditions from vacuum stability and triviality 
in the three-loop radiative seesaw model with TeV-scale right-handed 
neutrinos which are odd under the $Z_2$ parity. 
In this model, some of the neutrino Yukawa coupling constants 
can be of the order of one.
Requirement of strongly first order phase transition  
for successful electroweak baryogenesis also prefers 
order-one coupling constants in the scalar sector.
Hence, it is important to clarify whether this model satisfies those 
theoretical conditions up to a given cutoff scale. 
It is found that the model can be consistent up to the scale above 10 TeV  
in the parameter region where the neutrino data, 
the lepton flavor violation data,  
the thermal relic abundance of dark matter 
as well as the requirement from the strongly first order phase transition 
are satisfied.

\end{abstract}

\maketitle

\setcounter{footnote}{0}
\renewcommand{\thefootnote}{\arabic{footnote}}

\section{Introduction}
 
 The Higgs sector is the last unknown part in the standard model (SM). 
 There is no compelling reason that it takes its minimal form. 
 Instead, there can be various possibilities for non-minimal Higgs sectors. 
 Currently, searches for the Higgs boson is underway at the Tevatron and the Large Hadron Collider (LHC). 
 It is expected that crucial information for the Higgs sector will be obtained in near future.
 On the other hand, there are phenomena which cannot explain within the SM such as 
 neutrino oscillation, existence of dark matter, and baryon asymmetry of the Universe. 
 They provide us strong motivations to consider physics beyond the SM. 
 
 The Higgs sector may be related to the physics to cause these phenomena.
 First of all, based on the WIMP (Weakly Interacting Massive Particle) hypothesis, the observed 
 value of the relic abundance requires the mass of the dark matter to be around the electroweak scale. 
 Why is the mass of the WIMP dark matter at the electroweak scale? This would strongly indicate
 a direct connection between the WIMP dark matter and the Higgs sector~\cite{Bento:2000ah,higgsportal}.
 Second, the origin of neutrino masses may also be the TeV scale: i.e., the origin of the 
 lepton number violation would be the existence of TeV-scale right-handed neutrinos or that 
 of lepton number violating scalar interactions in the extended Higgs sector.  
 As a natural scenario for generating tiny neutrino masses, such lepton number violation  
 at the TeV scale would be transmitted to the left-handed neutrino sector via loop effects of 
 the Higgs sectors. This is so-called the radiative seesaw scenario~\cite{zee,zeebabu,knt,ma,aks-prl}.
 Finally, a simple description of baryon asymmetry of the Universe may be the scenario in which 
 electroweak phase transition is of strongly first order in order to satisfy the 
 Sakharov's condition of departure from thermal equilibrium~\cite{sakharov}. 
This is the scenario of  electroweak baryogenesis~\cite{ewbg}.  

 An attractive point for these scenarios would be that they can give definite predictions to 
 TeV-scale phenomenology, so that we can directly test them by experiments, in principle.     
 Various scenarios for the WIMP dark matter can be tested by direct and indirect searches 
 as well as collider experiments~\cite{Bento:2000ah,higgsportal,Ma-DM}.
 Each radiative seesaw model predicts a characteristic extended Higgs sector with  
 lepton number violating interactions~\cite{zee,kkloy,zeebabu,babu-macesanu,szb} 
 or right-handed neutrinos~\cite{knt,kingman-seto,ma,radiative,CaoMa}. 
 It is known that the dynamics for the electroweak baryogenesis strongly 
 related to the Higgs boson phenomenology at colliders~\cite{ewbg-thdm2}.

 Therefore, an interesting question is whether or not we can construct 
 a {\it successful} 
 TeV-scale phenomenological model where these three phenomena would be explained simultaneously. 
 Recently a TeV-scale model has been proposed for such a purpose~\cite{aks-prl},  
 in which 1) tiny neutrino masses are generated without excessive fine tuning 
 at the three-loop level by the dynamics of 
 an extended Higgs sector and right-handed neutrinos under an unbroken $Z_2$ parity, 2)    
 the $Z_2$ parity also guaranties the stability of a dark matter candidate which is a $Z_2$ odd 
 scalar boson, and 3) the strongly first order phase transition for 
 successful electroweak baryogenesis can be realized by the non-decoupling effect 
 in the Higgs sector.  
 Phenomenology of this model has been discussed in Ref.~\cite{aks-prd}, and the 
 related collider physics~\cite{typeX,stefano,ak-majorana} and dark matter properties~\cite{aks-cdms2} 
 have also been studied.  In these papers, phenomenologically allowed parameter regions have been mainly discussed. 
 
 In this paper, we investigate the theoretical constraint on the parameter regions in this model~\cite{aks-prl}  
 from the requirement of vacuum stability and perturbativity up to a given cutoff scale $\Lambda$ of the model~\cite{cabibbo}.
 In the present  model, there is no mechanism for cancellation of the quadratic divergences which appear in the 
 renormalization calculation for the Higgs boson mass, so that a huge fine tuning is required 
 if $\Lambda$ is much higher than the electroweak scale.  
 To avoid such an unnatural situation, we need to consider $\Lambda$ to be at most ${\cal O}(10)$ TeV, 
 above which the model would be replaced by a more fundamental theory~\cite{Murayama}. 
 Hence, we have to study the theoretical consistency of the model up to such values for $\Lambda$. 
 In particular, some of the neutrino Yukawa coupling constants are of order one in magnitude, as 
 the scale of tiny neutrino masses is generated by loop dynamics so that we do not need fine tuning 
 for the size of the coupling constants.  In addition, some of the coupling constants in the Higgs potential 
 are of order one to realize the non-decoupling one-loop effect for strongly first order phase transition. 
 Although the parameters discussed in the previous works satisfy the bound from tree-level unitarity~\cite{pu_nonminimum}, 
 it is non-trivial that the model can be consistent at the quantum level 
 with the theoretical requirements up to $\Lambda \sim 10$ TeV. 
 Theoretical bounds from vacuum stability and perturbativity have been used to constrain parameters 
 in extended Higgs sectors such as the two Higgs doublet model~\cite{vs_thdm} and the Zee model~\cite{kkloy}.
 Here we apply the similar analysis to the model.
 We prepare a full set of the renormalization group equations (RGEs) for dimensionless coupling constants  
 in the model at the one-loop level, and analyze the behavior of running coupling constants. 
 
 We also revise the phenomenological constraint from lepton flavor violation (LFV) in the model. 
 In the previous analysis~\cite{aks-prl,aks-prd} only 
 the constraint from $\mu\to e\gamma$ data has been taken into account. 
 Here, we also analyze the one-loop induced $\mu\to eee$ process, whose current experimental data~\cite{meee} 
 turn out to give a stronger bound on the parameter 
 space than those of $\mu \to e \gamma$~\cite{MEGA}\footnote{
 We thank Toru Goto for telling us the possibility of a sizable contribution to the branching ratio of $\mu\to eee$ in our model.}.
 
The paper is organized as follows. 
 In Sec.II, we give a brief review of the model. 
In Sec.III, we introduce the criteria for vacuum stability and triviality, and  
  numerical results of the RGE analysis are presented.  
In Sec.IV, we give a conclusion.

\section{The three-loop neutrino mass model}

We give a short review for the the three-loop neutrino mass model 
in Ref.~\cite{aks-prl} to make the paper self-contained.  

\subsection{Model}
In this model, two Higgs doublets ($\Phi_1$ and $\Phi_2$) with hypercharge 
$Y=1/2$, charged scalar singlets ($S^\pm$), a real scalar singlet 
($\eta$) and right-handed neutrinos ($N_R^\alpha$ with $\alpha =1,2$) 
are introduced.
We impose two kinds of discrete symmetries; i.e., 
$Z_2$ and $\tilde{Z}_2$ to the model. 
The former, which is exact, is introduced in order to  
forbid the tree-level Dirac neutrino mass term 
and at the same time to guarantee the stability of dark matter. 
The latter one, which is softly broken, is introduced to avoid 
the tree-level flavor changing neutral current~\cite{fcnc}. 
Under the $\tilde{Z}_2$ symmetry there are four types of Yukawa 
interactions~\cite{Ref:Barger,grossman}. In our model~\cite{aks-prl}, so-called the 
type-X Yukawa interaction~\cite{typeX,typeX2} is favored since the charged Higgs boson 
from the two doublets can be taken to be as light as around 100 GeV 
without contradicting the $b\to s \gamma$ data. 
Such a light charged Higgs boson is important to reproduce 
the correct magnitude of neutrino masses. 
The particle properties under the discrete symmetries are shown 
in Table~\ref{z2},
where $Q_L^i$, $u_R^i$, $d_R^i$, $L_L^i$ and $e_R^i$ are the $i$-th 
generation of the left-handed quark doublet, the 
right-handed up-type quark singlet, the 
left-handed lepton doublet and the right-handed charged lepton singlet, 
respectively.
\begin{table}[t]
\begin{center}
\begin{tabular}{|c||c|c|c|}\hline
&$Q_L^i\hspace{1mm}u_R^{i}\hspace{1mm}d_R^{i}\hspace{1mm}L_L^i \hspace{1mm}e_R^{i}$&$\Phi_1\hspace{1mm}\Phi_2$&$S^{\pm}\hspace{1mm}\eta \hspace{1mm}N_R^\alpha$\\\hline\hline
$Z_2$(exact)&$+\hspace{2mm}+\hspace{2mm}+\hspace{2mm}+\hspace{2mm}+$&$+\hspace{2mm}+$&$-\hspace{2mm}-\hspace{2mm}-$\\\hline
$\tilde{Z}_2$(softly broken)&$+\hspace{2mm}-\hspace{2mm}-\hspace{2mm}+\hspace{2mm}+$&$+\hspace{2mm}-$&$+\hspace{2mm}-\hspace{2mm}+$\\\hline
\end{tabular}
\caption{Particle properties under the discrete symmetries.}
\label{z2}
\end{center}
\end{table}

The type-X Yukawa interaction is given by
\begin{align}
\mathcal{L}_{\text{yukawa}}^{\text{Type-X}}&=-\sum_{i,j}
\left[\left(\bar{Q}_L^i Y^{d}_{ij} \Phi_2 d_R^j \right) 
+ \left(\bar{Q}_L^i Y^{u}_{ij} \Phi^c_2u_R^j \right)
+ \left(\bar{L}_L^{i} Y^{e}_{ij} \Phi_1 e_R^j \right)\right]+{\rm h.c.},   \label{yukawa1}
\end{align}
where Yukawa coupling matrix for leptons is diagonal, 
$Y^e_{ij}={\rm diag}(y_{e^1}, y_{e^2}, y_{e^3})$.
The mass term and the Yukawa interaction for $N_R^\alpha$ are written as 
\begin{align}
\mathcal{L}_{N_R}&=\sum_{\alpha=1}^2
\frac{1}{2}m_{N_R^\alpha}
\overline{(N_R^{\alpha})^c}N_R^\alpha-\sum_{i=1}^{3}\sum_{\alpha=1}^2
\left[h_i^\alpha\overline{(e_R^i)^c}N_R^\alpha S^++{\rm h.c.}\right].   \label{yukawa2}
\end{align}
The scalar potential is given by
\begin{align}
V=&+\mu_1^2|\Phi_1|^2+\mu_2^2|\Phi_2|^2
-\Big[\mu_3^2 \Phi_1^\dagger\Phi_2+ {\rm h.c.} \Big]\notag\\
& +\frac{1}{2}\lambda_1|\Phi_1|^4+\frac{1}{2}
\lambda_2|\Phi_2|^4+\lambda_3|\Phi_1|^2|\Phi_2|^2
+\lambda_4|\Phi_1^\dagger\Phi_2|^2
+\frac{1}{2}\Big[\lambda_5 (\Phi_1^\dagger\Phi_2)^2 
           +  {\rm h.c.} \Big] \notag\\
&+\mu_S^2|S^-|^2+\rho_1|S^-|^2|\Phi_1|^2+\rho_2|S^-|^2|\Phi_2|^2+\frac{1}{4}\lambda_S|S^-|^4\notag\\
&+\frac{1}{2}\mu_\eta^2\eta^2+\frac{1}{2}\sigma_1\eta^2|\Phi_1|^2+\frac{1}{2}\sigma_2\eta^2|\Phi_2|^2+\frac{1}{4!}\lambda_\eta\eta^4\notag\\
&+\sum_{a,b=1}^2\Big[\kappa\epsilon_{ab}(\Phi_a^c)^\dagger\Phi_b S^-\eta+{\rm h.c.}\Big]+\frac{1}{2}\xi|S^-|^2\eta^2,  \label{pot}
\end{align}
where $\epsilon_{ab}$ are anti-symmetric matrices with $\epsilon_{12}=1$. 
The parameters $\mu_3^2$, $\lambda_5$, and $\kappa$ are complex numbers. 
Two of their phases can be absorbed by rephasing the fields, and the 
rest is a physical one. In this paper, we neglect this CP-violating 
phase for simplicity. 
The Higgs doublets are parameterized as
\begin{align}
\Phi_i=\left(\begin{array}{c}
w_i^+\\
\frac{1}{\sqrt{2}}(h_i+v_i+iz_i)
\end{array}\right),
\end{align}
where $v_i$ are vacuum expectation values (VEVs) of the Higgs fields,  
and these are constrained by $v (=\sqrt{v_1^2+v_2^2}) \simeq 246$ GeV. 
The ratio of the two VEVs is defined by $\tan\beta =v_2/v_1$. 
The physical scalar states $h$, $H$, $A$ and $H^\pm$ 
in the $Z_2$ even sector can be obtained mixing angles $\alpha$ and $\beta$,
\begin{align}
\left(\begin{array}{c}
w_1^\pm\\
w_2^\pm
\end{array}\right)
=R(\beta)\left(\begin{array}{c}
w^\pm\\
H^\pm
\end{array}\right),\hspace{3mm}\left(\begin{array}{c}
z_1\\
z_2
\end{array}\right)
=R(\beta)\left(\begin{array}{c}
z\\
A
\end{array}\right),\hspace{3mm}
\left(\begin{array}{c}
h_1\\
h_2
\end{array}\right)
=R(\alpha)\left(\begin{array}{c}
H\\
h
\end{array}\right),
\end{align}
where $w^\pm$ and $z$ are the Nambu-Goldstone bosons absorbed 
by the longitudinal weak gauge bosons, and 
the rotation matrix with the angle $\theta$ is given  by 
\begin{align}
R(\theta)=
\left(\begin{array}{cc}
\cos\theta & -\sin\theta\\
\sin\theta & \cos\theta
\end{array}\right).
\end{align}
The mass formulae of physical scalar states are given by
\begin{align}
m_A^2&=M^2-v^2\lambda_5, \label{mA}\\
m_{H^\pm}^2&=M^2-\frac{v^2}{2}(\lambda_4+\lambda_5),\label{mHpm}\\
m_H^2&=\cos^2(\alpha-\beta) M_{11}^2+2\sin(\alpha-\beta)\cos(\alpha-\beta) M_{12}^2+\sin^2(\alpha-\beta) M_{22}^2,\\
m_h^2&=\sin^2(\alpha-\beta) M_{11}^2-2\sin(\alpha-\beta)\cos(\alpha-\beta) M_{12}^2+\cos^2(\alpha-\beta) M_{22}^2,\\
m_{S}^2&=\mu_S^2+\frac{v^2}{2}\rho_1\cos^2\beta+\frac{v^2}{2}\rho_2\sin^2\beta, \label{mS}\\
m_\eta^2&=\mu_\eta^2+\frac{v^2}{2}\sigma_1\cos^2\beta+\frac{v^2}{2}\sigma_2\sin^2\beta,
\end{align}
where $M (=\mu_3/\sqrt{\sin\beta\cos\beta})$ is the soft breaking scale 
for the $\tilde{Z}_2$ symmetry, and 
\begin{align}
M_{11}^2&=v^2(\lambda_1\cos^4\beta+\lambda_2\sin^4\beta)+\frac{v^2}{2}\lambda\sin^22\beta,\\
M_{22}^2&=M^2+v^2\sin^2\beta\cos^2\beta(\lambda_1+\lambda_2-2\lambda),\\
M_{12}^2&=v^2\sin\beta\cos\beta(-\lambda_1\cos^2\beta+\lambda_2\sin^2\beta+\lambda\cos2\beta),
\end{align} 
with $\lambda=\lambda_3+\lambda_4+\lambda_5$. 
Notice that $M$, $\mu_S$ and $\mu_\eta$ are free mass parameters 
irrelevant to the electroweak symmetry breaking.

\subsection{Neutrino mass and mixing}

\begin{figure}[t]
\begin{center}
  \epsfig{file=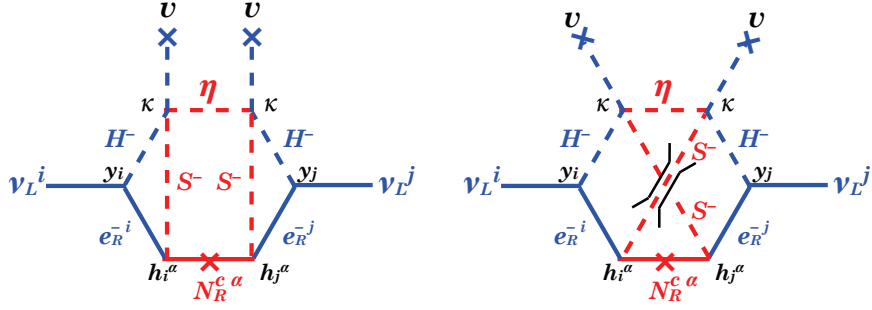,width=12cm}
\end{center}
  \caption{The Feynman diagrams for generating tiny neutrino masses. }
  \label{diag-numass}
\end{figure}

The neutrino mass matrix $M^\nu_{ij}$ is generated by the three-loop diagrams in
FIG.~\ref{diag-numass}.
The absence of lower order loop contributions is guaranteed by the 
$Z_2$ symmetry.  
The resulting mass matrix is calculated as 
\begin{eqnarray}
M^\nu_{ij} &=& \sum_{\alpha=1}^2 
      4 \kappa^2 \tan^2\beta 
  (y_{\ell_i}^{\rm SM} h_i^\alpha) (y_{\ell_j}^{\rm SM} h_j^\alpha)
   F(m_{H^\pm}^{},m_{S}^{},m_{N_R^{\alpha}}, m_\eta), \label{eq:mij}
\end{eqnarray}
where the loop integral function $F$ is given by
\begin{eqnarray}
&&F(m_{H^\pm}^{},m_S^{},m_{N}, m_\eta) =
   \left(\frac{1}{16\pi^2}\right)^3 \frac{(-m_N^{})}{m_N^2-m_\eta^2}
 \frac{v^2}{m_{H^\pm}^4}\nonumber\\
 && \times \int_0^{\infty} x dx
  \left\{B_1(-x,m_{H^\pm}^{},m_S^{})-B_1(-x,0,m_S^{})\right\}^2
  \left(\frac{m_N^2}{x+m_N^2}-\frac{m_\eta^2}{x+m_\eta^2}\right),
\end{eqnarray}
with $y_{\ell_i}^{\rm SM}=\sqrt{2}m_{\ell_i}/v$, where 
$\ell_1$, $\ell_2$ and $\ell_3$ correspond to 
$e$, $\mu$ and $\tau$, respectively. 
The function $B_1$ is the tensor coefficient
in the formalism by Passarino-Veltman 
for one-loop integrals~\cite{passarino-veltman}.
In the following discussion, we take $m_{N_R^1}=m_{N_R^2}\equiv m_{N_R}$, 
for simplicity. 
Numerically, the magnitude of the function $F$ is of order $10^{4}$ eV
in the wide range of parameter regions of our interest.  
Since $y_{\ell_i}^{\rm SM} < 10^{-2}$, the correct 
scale of neutrino masses can be naturally obtained from 
the three-loop diagrams.

The generated mass matrix 
$M^\nu_{ij}$ in Eq.~(\ref{eq:mij}) of neutrinos
can be related to the neutrino oscillation data by
\begin{eqnarray}
 M^\nu_{ij} = U_{is} (M^\nu_{\rm diag})_{st} (U^T)_{tj}, 
\end{eqnarray}
where $M^\nu_{\rm diag}$ $=$ ${\rm diag}(m_1, m_2, m_3)$.
For the case of the normal hierarchy we identify the mass eigenvalues
as  
$m_1=0$, $m_2=\sqrt{\Delta m_{\rm solar}^2}$, $m_3=\sqrt{\Delta m_{\rm
atm}^2}$, while  for inverted hierarchy $m_1=\sqrt{\Delta m_{\rm atm}^2}$,
$m_2=\sqrt{\Delta m_{\rm atm}^2 + \Delta m_{\rm solar}^2}$ and $m_3=0$
are taken.
The Maki-Nakagawa-Sakata matrix $U_{is}$~\cite{mns} is parameterized as 
\begin{eqnarray}
  U=\left[\begin{array}{ccc}
     1&0&0\\
     0&c_{23}^{}&s_{23}^{}\\
     0&-s_{23}^{}&c_{23}^{}\\
           \end{array}
     \right]
  \left[\begin{array}{ccc}
     c_{13}^{}&0&s_{13}^{}e^{i\delta}\\
     0& 1 &0\\
     -s_{13}^{}e^{-i\delta}&0&c_{13}^{}\\
           \end{array}
     \right]
  \left[\begin{array}{ccc}
     c_{12}^{}&s_{12}^{}&0\\
     -s_{12}^{}&c_{12}^{}&0\\
         0&0&1
          \end{array}
     \right]
    \left[\begin{array}{ccc}
     1&0&0\\
     0&e^{i\tilde{\alpha}}&0\\
         0&0&e^{i\tilde{\beta}}
          \end{array}
     \right],
\end{eqnarray}
where $s_{ij}$ and $c_{ij}$ represent $\sin\theta_{ij}$ and
$\cos\theta_{ij}$, respectively, with $\theta_{ij}$ to be the neutrino mixing angle 
between the $i$th and $j$th generations, and $\delta$ is the Dirac phase while
$\tilde{\alpha}$ and $\tilde{\beta}$ are Majorana phases.
For simplicity, we neglect the effects of these CP violating 
phases in the following 
analysis.
Current neutrino oscillation data give the following 
values~\cite{pdg};  
\begin{eqnarray}
\Delta m^2_{\rm solar} \simeq 7.59\times 10^{-5} \,{\rm eV}^2 \,,~~
|\Delta m^2_{\rm atm}| \simeq 2.43\times 10^{-3} \,{\rm eV}^2\,, \\
\sin^2\theta_{12} \simeq 0.32 \,,~~~~
\sin^2\theta_{23} \simeq 0.5 \,,~~~~
\sin^2\theta_{13} < 0.04\,.~~~~~~
\label{obs_para}
\end{eqnarray}

In the next subsection, we discuss parameter regions in which 
both neutrino data and the LFV data are satisfied.

  \subsection{Lepton flavor violation}
  
\begin{figure}[t]
\begin{center}
  \epsfig{file=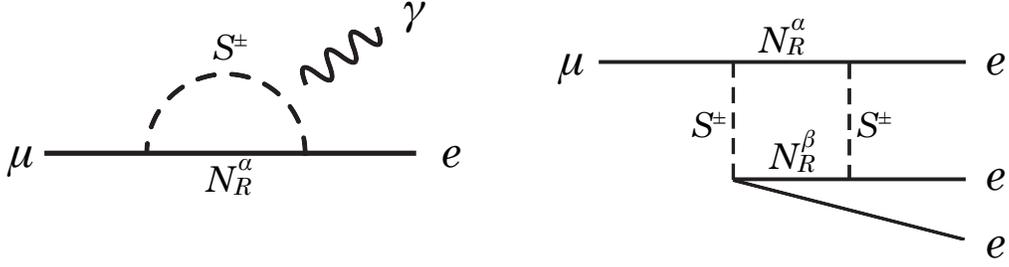,width=14cm}
\end{center}
  \caption{The LFV processes.}
  \label{diagram}
\end{figure}

The model receives the severe constraints from the lepton-flavor violating processes of 
$\mu \to e \gamma$ and $\mu\to eee$: see Fig.~\ref{diagram}.
These processes are induced through one-loop diagrams by
$N_R^\alpha$  and $S^\pm$ with the Yukawa couplings $h_i^\alpha$
($i=e$ and $\mu$).
The branching ratio of $\mu \to e \gamma$ is given by
\footnote{The formula of Eq.(\ref{mer}) is different from Eq.(31) in Ref.~\cite{aks-prd}  which includes errors.
We have recalculated the values of $B(\mu\to e\gamma)$ 
by using the corrected formula and checked 
that the values of $B(\mu\to e\gamma)$ in the parameter sets in Table~II 
in Ref.~\cite{aks-prd} are still below the experimental bound.}
\begin{equation}
B(\mu\to e \gamma) \simeq  \frac{3\alpha_{\rm em} v^4}{32\pi m_{S}^4}\left|
\sum_{\alpha=1}^2 h_e^{\alpha\ast} h_\mu^\alpha F_2\left(\frac{m_{N_R^\alpha}^2}{m_{S}^2}\right)
\right|^2 ,
\label{mer}
\end{equation}
where $F_2(x)\equiv (1-6x+3x^2+2x^3-6x^2\ln x)/6(1-x)^4$.
For $\mu \to eee$, the branching ratio is calculated by 
\begin{align}
&B(\mu \to eee)=\frac{1}{16 G_F^2}\left(\frac{1}{16\pi^2}\right)^2\Bigg|\sum_{\alpha,\beta=1}^2(h_\mu^{\alpha *} h_e^{\alpha *} h_e^\beta h_e^\beta) 
\left[\frac{m_{N_R^\alpha} m_{N_R^\beta}}{(m_{N_R^\alpha}^2-m_{S}^2)(m_{N_R^\beta}^2-m_{S}^2)}\right]\notag\\
&\times \left(-\frac{m_{N_R^\alpha}^2+m_{N_R^\beta}^2}{m_{N_R^\alpha}^2-m_{N_R^\beta}^2}\log\frac{m_{N_R^\beta}}{m_{N_R^\alpha}}+\frac{m_{N_R^\alpha}^2+m_{S}^2}{m_{N_R^\alpha}^2-m_{S}^2}\log\frac{m_{S}}{m_{N_R^\alpha}}
+\frac{m_{N_R^\beta}^2+m_{S}^2}{m_{N_R^\beta}^2-m_{S}^2}\log\frac{m_{S}}{m_{N_R^\beta}}
+1\right)\Bigg|^2. \label{meee1}
\end{align}
In particular, when $m_{N_R^1}=m_{N_R^2}$ $(=m_{N_R})$, the expression in Eq.~(\ref{meee1}) is reduced to  
\begin{align}
B(\mu \to eee)=&\frac{1}{4G_F^2}\left(\frac{1}{16\pi^2}\right)^2\left|\sum_{\alpha,\beta=1}^2h_\mu^{\alpha *} h_e^{\alpha *} h_e^\beta h_e^\beta \right|^2 \notag\\
&\times
\left(\frac{m_{N_R}}{m_{N_R}^2-m_{S}^2}\right)^4\left(\frac{m_{N_R}^2+m_{S}^2}{m_{N_R}^2-m_{S}^2}\log\frac{m_{S}}{m_{N_R}}+1\right)^2. \label{meee2}
\end{align}

Assuming that $h_e^\alpha \sim {\cal O}(1)$, 
the masses of $N_R^\alpha$ and $S^\pm$ are strongly constrained from below. 
In particular, if we assume that $m_{S}\gsim 400$ GeV, $m_{N_R}\gsim {\cal O}(1)$ TeV 
is required to satisfy the current experimental bounds, $B(\mu\to e \gamma)< 1.2 \times 10^{-11} $~\cite{MEGA} 
and $B(\mu\to eee)< 1.0 \times 10^{-12} $~\cite{meee}.
Such a relatively heavier $S^\pm$ is favored from the discussion on 
the dark matter relic abundance and electroweak baryogenesis~\cite{aks-prl,aks-prd}.

\subsection{Typical scenarios}

In Table~\ref{h-numass}, we show four choices 
for the parameter sets, and resulting values for   
the neutrino Yukawa coupling constants $h_i^\alpha$ which satisfy 
the neutrino data and the LFV data. 
For all parameter sets, $m_S=400$ GeV and $m_{N_R}=5$ TeV are assumed.
Set A and Set B are taken as the normal hierarchy in the neutrino masses with
$\sin^2\theta_{13}=0$ and 0.03, respectively, while 
Set C and Set D are for the inverted hierarchy. 
The predictions on $B(\mu\to e\gamma)$ and $B(\mu \to eee)$ are also shown in the table\footnote{
In Table~\ref{h-numass}, 
we show the numbers of the $h_i^\alpha$ coupling constants 
with four digits for Set C,   
because the branching ratios of $\mu\to e\gamma$ 
and $\mu\to eee$ are sensitive to these numbers 
due to large cancellations.}.
The scenario with the inverted hierarchy requires the larger values for
$\kappa \tan\beta$, so that the normal hierarchy scenarios are more natural in our model.  
 \begin{table}[t]
\begin{center}
  \begin{tabular}{|c||c|c||c|c|c|c|c|c||c|c|}\hline
   \mbox{Set} & \multicolumn{2}{c||}{\mbox{Inputs}}   
    & \multicolumn{6}{c||}{\mbox{Yukawa~couplings}} &
 \multicolumn{2}{c|}{\mbox{LFV}}  \\ \hline
 &  $\sin^2\theta_{13}$  & $\kappa\tan\beta$ & $h_e^1$ & $h_e^2$ & $h_\mu^1$ & $h_\mu^2$ & $h_\tau^1$ & $h_\tau^2$  &
   $B(\mu\!\!\to\!\! e\gamma)$ &$B(\mu\!\!\to\!\! 3e)$
   \\\hline \hline
   A&0&  54 &  1.2 &  1.3 &   0.024  & -0.011 &  0.00071 &  -0.0014
   &$2.8\!\times \!10^{-14}$ &$5.9\!\times \!10^{-13}$ 
   \\ \hline 
  B& 0.03 &76
  & 1.1    &1.1   &0.0028  & 0.018
  &-0.00055& 0.00097
   &$6.1\!\times \!10^{-14}$ &$7.4\!\times \!10^{-13}$ 
   \\ \hline \hline 
  C& 0 & 80
  & 3.500    & 3.474   &0.01200  & -0.01192
  & -0.0007136 & 0.0007086
   &$4.4\!\times \!10^{-17}$ &$6.2\!\times \!10^{-14}$ 
   \\ \hline 
  D& 0.03 & 128
  & 2.1    & 2.2   &0.0064  & -0.0086
  & -0.00053 & 0.00035
   &$3.5\!\times \!10^{-15}$ &$7.0\!\times \!10^{-13}$ 
      \\ \hline 
   \end{tabular}
\end{center}
\vspace{-2mm}  \caption{Values of $h_i^\alpha$ as well as those of 
branching ratios of $\mu\to e \gamma$ and $\mu\to 3e$ 
for $m_\eta=50$ GeV and $m_{H^\pm}^{}=100$ GeV for various scenarios 
which satisfy neutrino data:     
Set A and Set B are scenarios of the the normal hierarchy 
while Set C and Set D are those of the inverted hierarchy. 
For all sets, $m_S=400$ GeV and $m_{N_R}=5$ TeV are taken.}
\label{h-numass}
\end{table}

\begin{figure}[t]
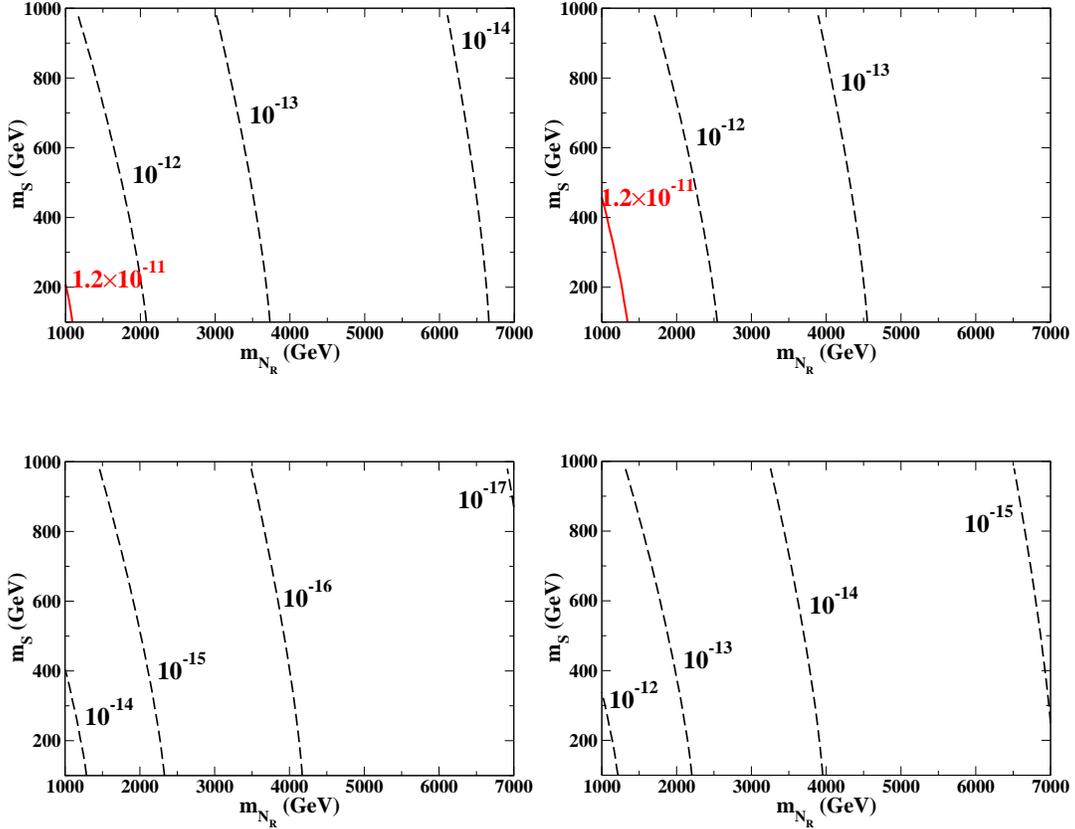

\begin{center}
  \epsfig{file=meg_A.eps,width=7cm}
  \epsfig{file=meg_B.eps,width=7cm}\\ \vspace{1cm}
  \epsfig{file=meg_C.eps,width=7cm}
  \epsfig{file=meg_D.eps,width=7cm}
\end{center}
  \caption{Contour plots for the branching ratio of $\mu\to e\gamma$ for the neutrino Yukawa coupling constants in 
           Set A (top-left), Set B (top-right), Set C (bottom-left) and Set D (bottom-right). 
           The contour for the upper limit from the data is given as the (red) solid curve.}  \label{LFV_1} 
\end{figure}

\begin{figure}[t]
\begin{center}
  \epsfig{file=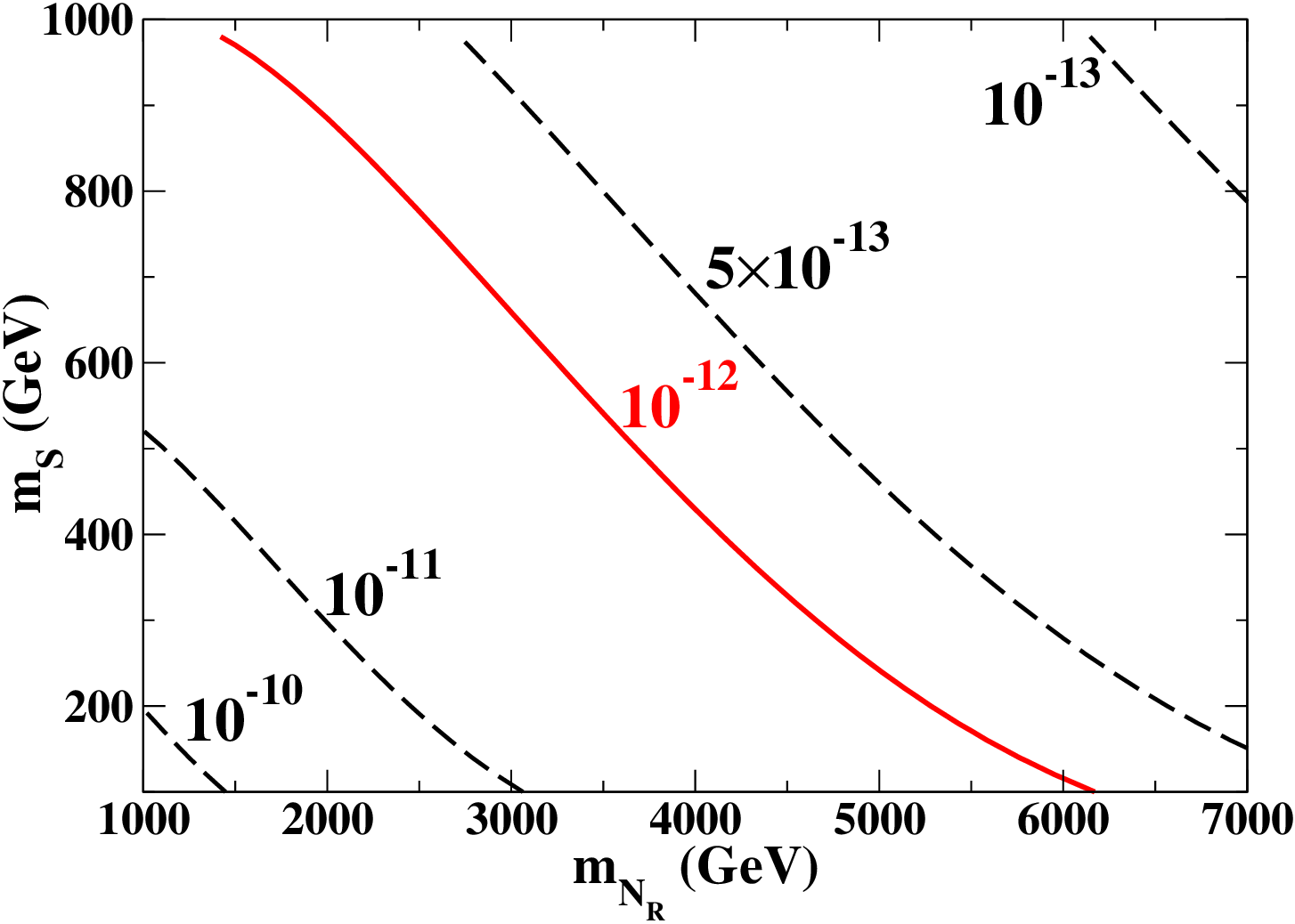,width=7cm}
  \epsfig{file=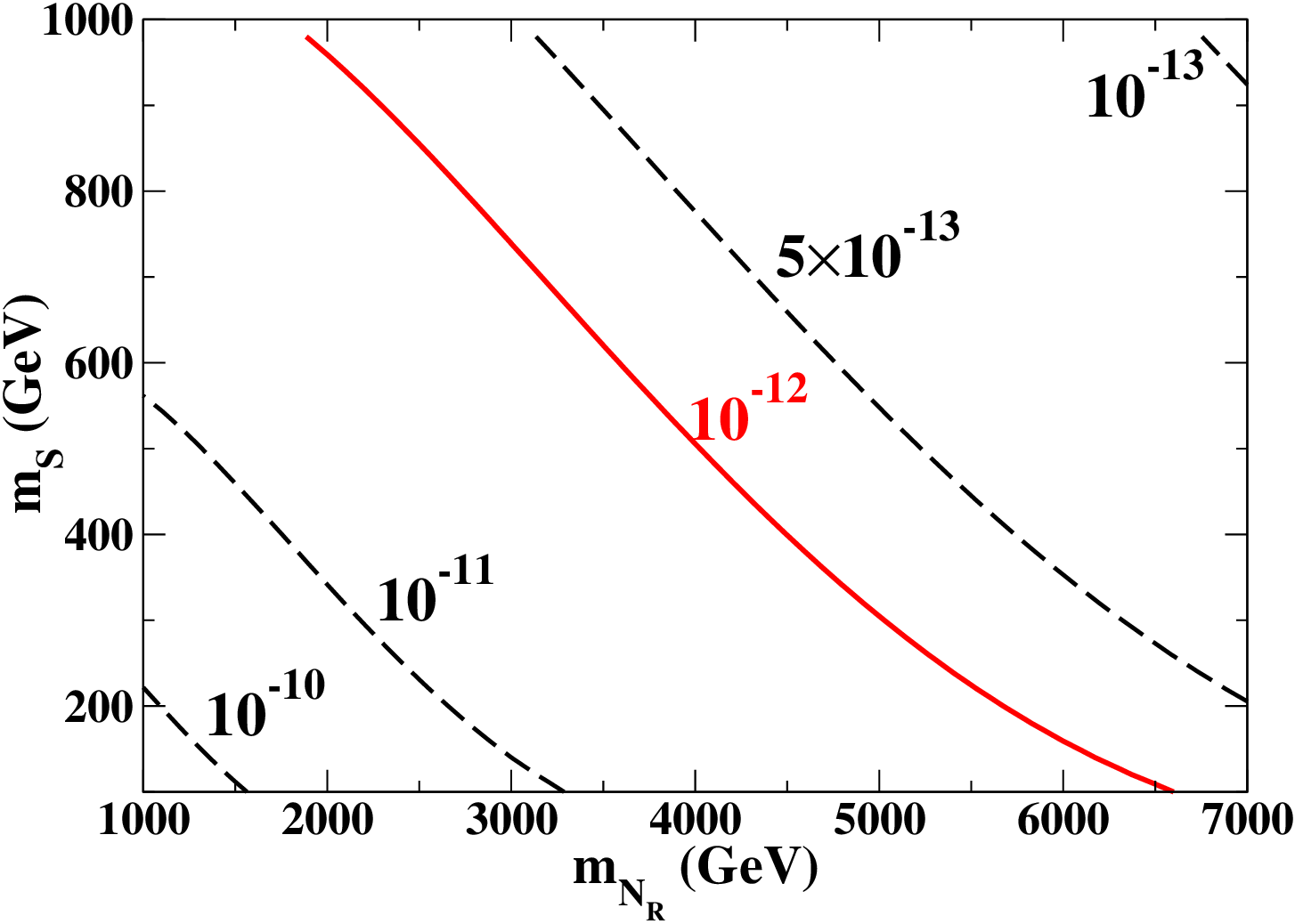,width=7cm}\\ \vspace{1cm}
  \epsfig{file=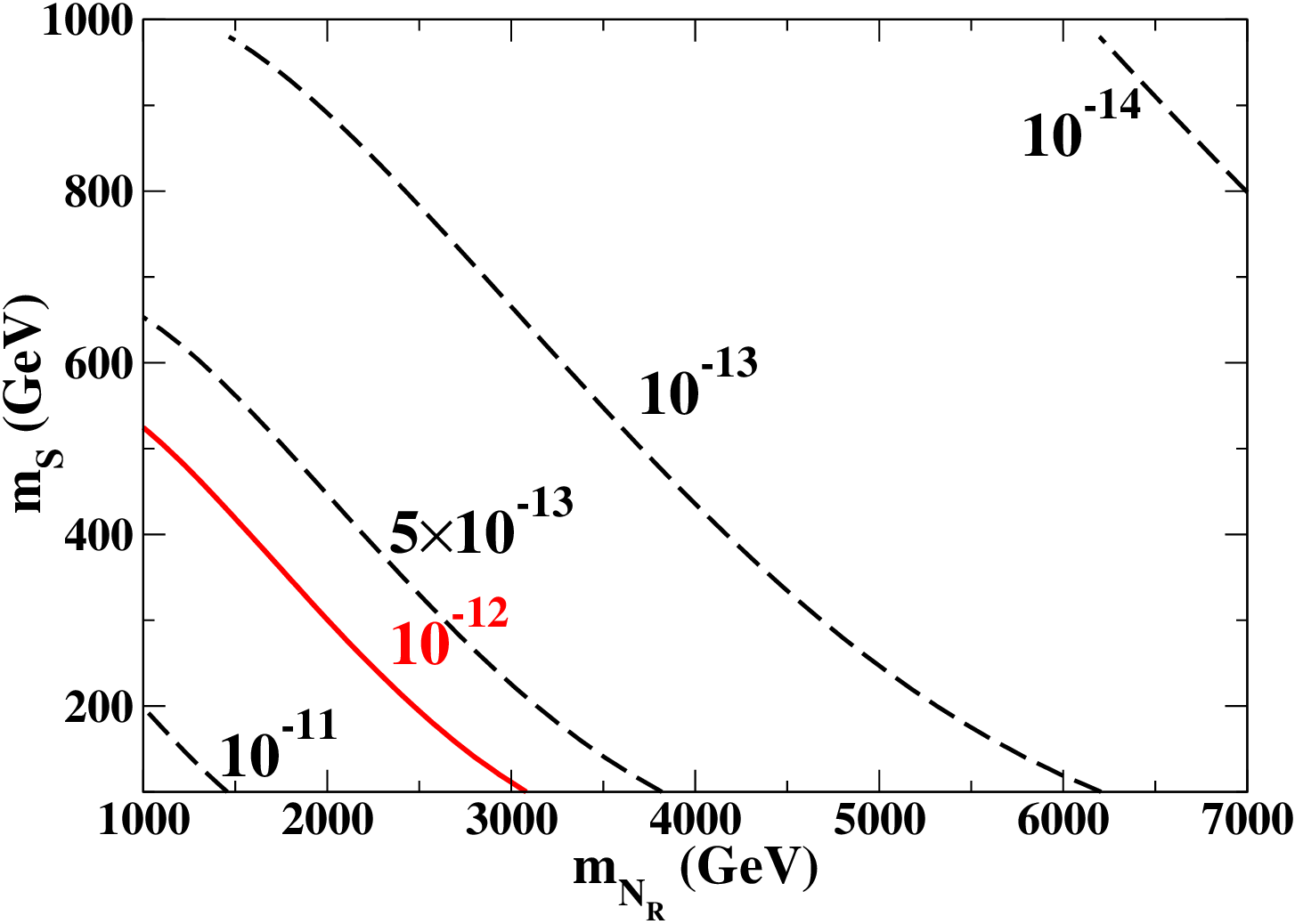,width=7cm}
  \epsfig{file=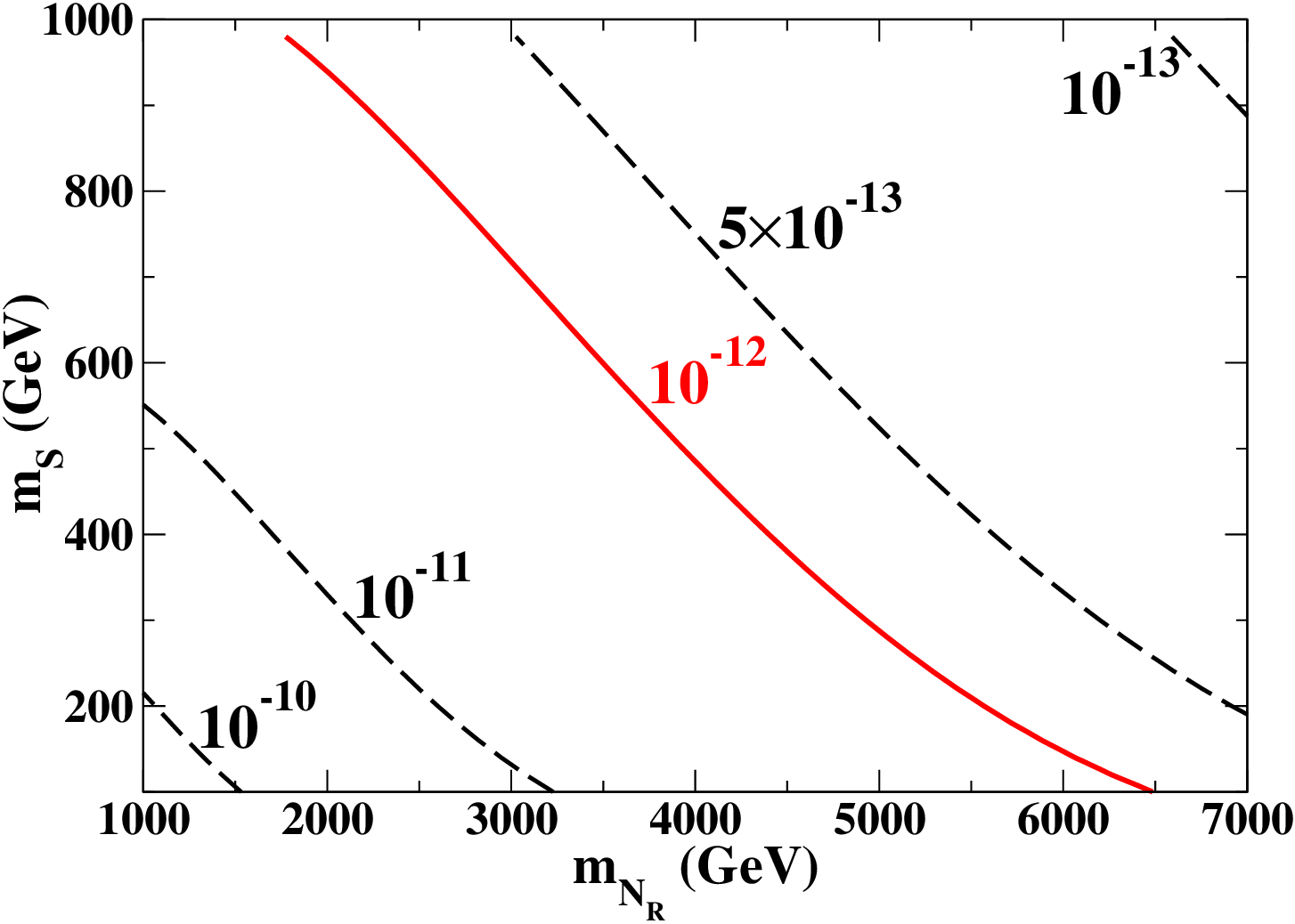,width=7cm}
\end{center}
  \caption{Contour plots for the branching ratio of $\mu\to eee$ for the neutrino Yukawa coupling constants in 
           Set A (top-left), Set B (top-right), Set C (bottom-left) and Set D (bottom-right). 
           The contour for the upper limit from the data is given as the (red) solid curve.}
  \label{LFV_2}
\end{figure}

In Fig.~\ref{LFV_1}, the contour plots of the branching ratio $B(\mu\to e\gamma)$ are shown in the $m_{S}^{}$-$m_{N_R}$ plane 
for the neutrino Yukawa coupling constants in Set A to Set D, 
while those of the branching ratio $B(\mu\to eee)$ are shown for these scenarios in Fig.~\ref{LFV_2}.  
The scale of the branching ratio of $\mu\to e\gamma$ is determined by $m_{N_R}$ and is insensitive to $m_{S}$, 
while that of $\mu\to eee$ largely depend on both $m_{N_R}$ and $m_{S}$ especially for Set A, Set B and Set D. 
It can easily be seen that a much stronger constraint comes from $\mu\to eee$ for all scenarios.


\subsection{Dark matter and electroweak phase transition}

From now on, we employ Set A in Table~\ref{h-numass} for further phenomenological analyses.
In this scenario, masses of $N_R^\alpha$ are at the multi-TeV scale, so that it may be natural that 
the rest $Z_2$-odd neutral field $\eta$ is the candidate of dark matter.  
Since $\eta$ is a singlet under the SM gauge group,
the interactions with $Z_2$-even particles are only through the Higgs coupling.
When $m_\eta < m_W$, the $\eta$ field predominantly annihilates into $b \bar{b}$ and $\tau^+\tau^-$ 
through $s$-channel Higgs boson ($h$ and $H$) mediations.  
Strong annihilation occurs at $m_\eta \simeq m_H^{}/2$
(and $m_\eta \simeq m_h/2$) due to the resonance of $H$ ($h$) mediation in the $s$-channel diagrams.
The pair annihilation into two photons through one-loop diagrams by $H^\pm$ and $S^\pm$ can also be important if $\kappa$ 
is of the order one.
The relic abundance becomes consistent with the data ($\Omega_{\rm DM} h^2 \sim 0.11$~\cite{wimp}) for $m_\eta \sim 50-60$ GeV, when we take $m_H^{}=m_{H^\pm}^{}\simeq 100$ GeV, $m_h^{}\simeq 120$ GeV,
$m_{S}^{}\gsim 400$ GeV with $\kappa=1.5$, $\sigma_1=0.05$, $\sigma_2=0.03$, and $\tan\beta=36$. 
In such scenario, the typical spin-independent cross section for the scattering of dark matter with a proton 
is of order of $10^{-8}$ pb which is within the reach of the direct search experiments such as superCDMS and XMASS. 

When $m_\eta < m_h/2$, the (SM-like) Higgs boson $h$ can decay into a dark matter pair $\eta\eta$. 
The branching ratio of $h \to \eta\eta$ is evaluated as $34$ \% ($22$ \%) for $m_h=120$ GeV and $m_\eta=48$ ($55$) GeV  
when $\sigma_1=0.05$, $\sigma_2=0.03$ and $\tan\beta > 10$. 
The invisible decay of $h$ can be tested at the LHC when $B(h \to \eta\eta) > 25$ \%~\cite{LHCinv}. 
At the ILC, it is expected that the branching ratio for the invisible decay 
of a few \% can be detected~\cite{Schumacher:2003ss}. Therefore, the invisible decay in this model can be tested at the collider experiments.

Our model~\cite{aks-prl}  satisfies the conditions for baryogenesis~\cite{sakharov}. 
The $B$ number violating interaction is the sphaleron interaction.  
The additional CP violating phases are in the Higgs sector and in the Yukawa interaction.
The condition of departure from thermal equilibrium can be realized by the strong first order electroweak phase transition,
which requires a large 
tri-linear coupling of the order parameter 
in the expression of the high temperature expansion~\cite{hte} where 
only the bosonic loop can contribute\footnote{We note that such a non-decoupling effect due to the bosonic loop 
can also affect the quantum correction to the triple Higgs boson coupling~\cite{ewbg-thdm2,Ref:KOSY}. Such 
a large correction to the Higgs self-coupling can be an important signature for successful 
electroweak baryogeneis at collider experiments.}.
In our model, there are many additional scalars running in the
loop so that the large coupling can be easily realized~\cite{ewbg-thdm2}.
The strong first order phase transition is possible for large $m_{S}^{}$ 
and/or $m_A^{}$ with the large non-decoupling effect:
{\it e.g.} 
$m_S^{} \sim 400$ GeV,
$m_A^{} \sim 100$ GeV, 
$M=100$ GeV and $\mu_S^{}=200$ GeV,
where $M$ and $\mu_S^{}$ are the 
invariant masses in Eq.(\ref{mA}) and Eq.(\ref{mS}), respectively. 
The result is not sensitive to $\tan\beta$.

\section{Bounds from triviality and vacuum stability}

There are scalar bosons in this model, so that quadratic divergences 
appear in the one-loop calculation for their masses. 
Because there is no mechanism by which such quadratic divergences 
are eliminated, enormous fine tuning is required to realize the 
renormalized Higgs boson mass being at the weak scale with a very high cutoff scale.   
Allowing the 1~\% fine tuning, the cutoff scale is at most $\Lambda \sim O(10)$ TeV, 
above which the theory would be replaced by a more fundamental one~\cite{Murayama}. 
Unless a mechanism of cancellation of the quadratic divergences such as supersymmetry is 
implemented, to avoid excessive fine tuning the model should be regarded as an effective 
theory, whose cutoff scale $\Lambda$ is between $m_{N_R^\alpha}$ and $O(10)$ TeV. 
We then need to confirm the theoretical consistency of the model up to $\Lambda$~\cite{cabibbo}.
We here evaluate bounds on the parameter space from vacuum stability and perturbativity, 
and examine whether 
the theoretically allowed parameter region is consistent with that by the experimental 
data discussed in the previous sections. 

We have to consider these two bounds seriously because of the following reasons. 
First, this model includes many scalar fields, $e.g.$, $h$, $H$, $A$, $H^\pm$, $S^\pm$ and $\eta$, 
so that the scale dependent dimensionless coupling constants would be drastically changed  
by the loop corrections due to the scalar bosons. 
Second, some of the Yukawa coupling constants for right-handed neutrinos are necessarily 
of order one for a radiative generation of the tiny mass scale of the neutrinos at the three-loop level. 
Finally, to realize the first order electroweak phase transition, some of the scalar self-coupling 
constants has to be as large as of order one.  

In order to evaluate the vacuum stability bound and the triviality bound,  
we estimate the scale dependences of the dimensionless coupling constants 
by using the RGEs at the one-loop level. 
We have calculated the one-loop beta functions for all the coupling constants in this model.   
The full set of the beta functions is listed in Appendix.  
We take into account the threshold effects in the calculation of the scale dependent coupling constants. 
In the scale below the mass of $S^\pm$, we treat the theory without $N_R$ and $S^\pm$. 
In the scale between the masses of the $S^\pm$ and $N_R$, we treat the theory without $N_R$. 
In the scale higher than the mass of the $N_R$, we treat the theory with full particle contents.
 
\subsection{The conditions}

In this model, there are scalar fields $\Phi_i$ ($i=1,2$), $S^\pm$ and $\eta$, which contain 
eleven degrees of freedom which would share the order parameter. 
The four of them are eliminated because of the $SU(2)_L\times U(1)_Y $ gauge symmetry. 
In the remining seven dimensional parameter space, we require that for any direction 
the potential is bounded from below with keeping positiveness~\cite{deshpande}. 
In the SM, this requirement is satisfied when the Higgs self-coupling constant is positive. 
In this model, we put the following conditions on the the dimensionless coupling constants:   
\begin{align}
\lambda_1(\mu)>0,\quad \lambda_2(\mu)>0,\quad \lambda_S(\mu)>0,\quad \lambda_\eta(\mu) >0, \label{eq:vsc1}
\end{align}
\begin{align}
&\sqrt{\lambda_1(\mu)\lambda_2(\mu)}+\lambda_3(\mu)+\text{MIN}[0,\quad (\lambda_4(\mu)+\lambda_5(\mu)),\quad (\lambda_4(\mu)-\lambda_5(\mu))]>0,\notag\\
&\sqrt{\lambda_1(\mu)\lambda_S(\mu)/2}+\rho_1(\mu)>0,\quad \sqrt{\lambda_1(\mu)\lambda_\eta(\mu)/3}+\sigma_1(\mu)>0,\quad \sqrt{\lambda_2(\mu)\lambda_S(\mu)/2}+\rho_2(\mu)>0,\notag\\
&\sqrt{\lambda_2(\mu)\lambda_\eta(\mu)/3}+\sigma_2(\mu)>0,\quad \sqrt{\lambda_S(\mu)\lambda_\eta(\mu)/6}+\xi(\mu)>0, \label{eq:vsc2}
\end{align} 
\begin{align}
&2\lambda_1(\mu)+2\lambda_2(\mu)+4\lambda_3(\mu)+4\rho_1(\mu)+4\rho_2(\mu)+\lambda_S(\mu)+4\sigma_1(\mu)+4\sigma_2(\mu)\notag\\
&+\frac{2}{3}\lambda_\eta(\mu)+4\xi(\mu)-16\sqrt{2}|\kappa(\mu)| >0.\label{vs_kappa}  
\end{align}
The conditions in Eqs.~(\ref{eq:vsc1}) and (\ref{eq:vsc2}) are obtained by the similar way as in 
Ref.~\cite{kkloy}, while the last condition in Eq.~(\ref{vs_kappa}) is derived such that 
the term with the coupling constant $\kappa$ in the potential satisfies the positivity condition 
for the direction where the VEVs of the fields $\Phi_1$, $\Phi_2$, $S^\pm$ and $\eta$ are 
a common value.

We require that all the dimensionless running coupling constants do not blow up below $\Lambda$. 
Since we discuss the model within the scale where the perturbation calculation remains reliable,  
we here require that the running coupling constants do not exceed some critical value. 
In this paper, we impose the following criterion in the coupling constants in the Higgs potential Eq.~(\ref{pot})
and the Yukawa interaction in Eqs.~(\ref{yukawa1}) and (\ref{yukawa2}):
\begin{align}
&|\lambda_i(\mu)|, \hspace{2mm}|\sigma_i(\mu)|, \hspace{2mm}|\rho_i(\mu)|, \hspace{2mm} |\kappa(\mu)|, \hspace{2mm} |\xi(\mu)| < 8\pi,\notag\\
&y_t^2(\mu),\hspace{2mm} y_b^2(\mu),\hspace{2mm}  y_\tau^2(\mu),\hspace{2mm}  (h_i^\alpha)^2(\mu)< 4\pi.  \label{vs3}
\end{align}
The similar critical value has been adopted in the analyses in the two Higgs dobulet model~\cite{vs_thdm} and in the Zee model~\cite{kkloy}.

\subsection{Allowed regions in the parameter space}

In this section, we evaluate allowed regions in parameter space,  
which satisfy the conditions of triviality and vacuum stability 
for each fixed cutoff scale $\Lambda$.
For the scenarios of the neutrino Yukawa coupling constants 
as well as the masses of right-handed neutrinos, we choose 
Set A in Table~\ref{h-numass}.  
We investigate the allowed regions in the $m_S$-$m_A$ plane, and 
the rest of the mass parameters in the scalar sector is fixed 
as shown in Teble~\ref{invariantmass}.

\begin{figure}[t]
\includegraphics[width=75mm]{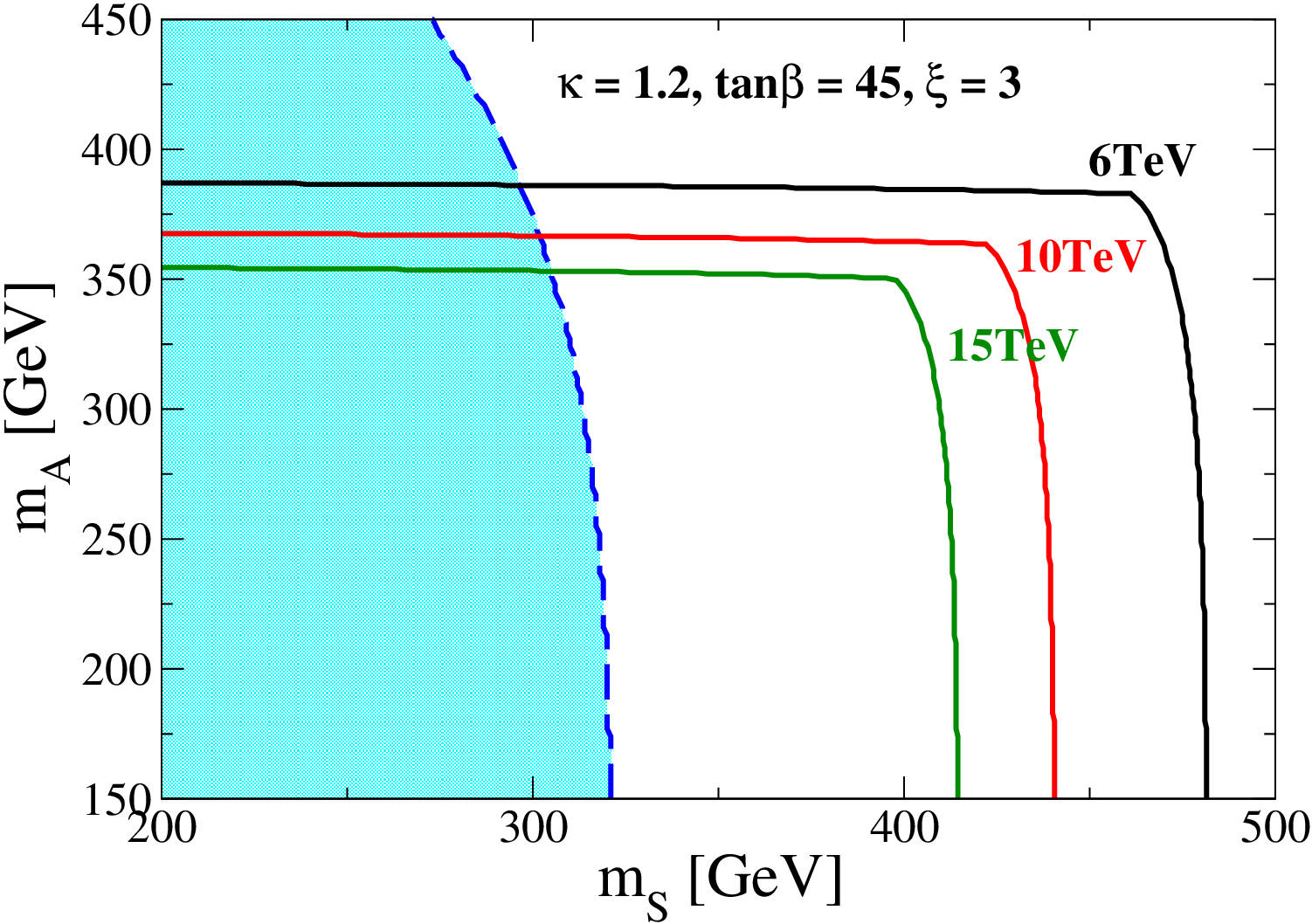}
\includegraphics[width=75mm]{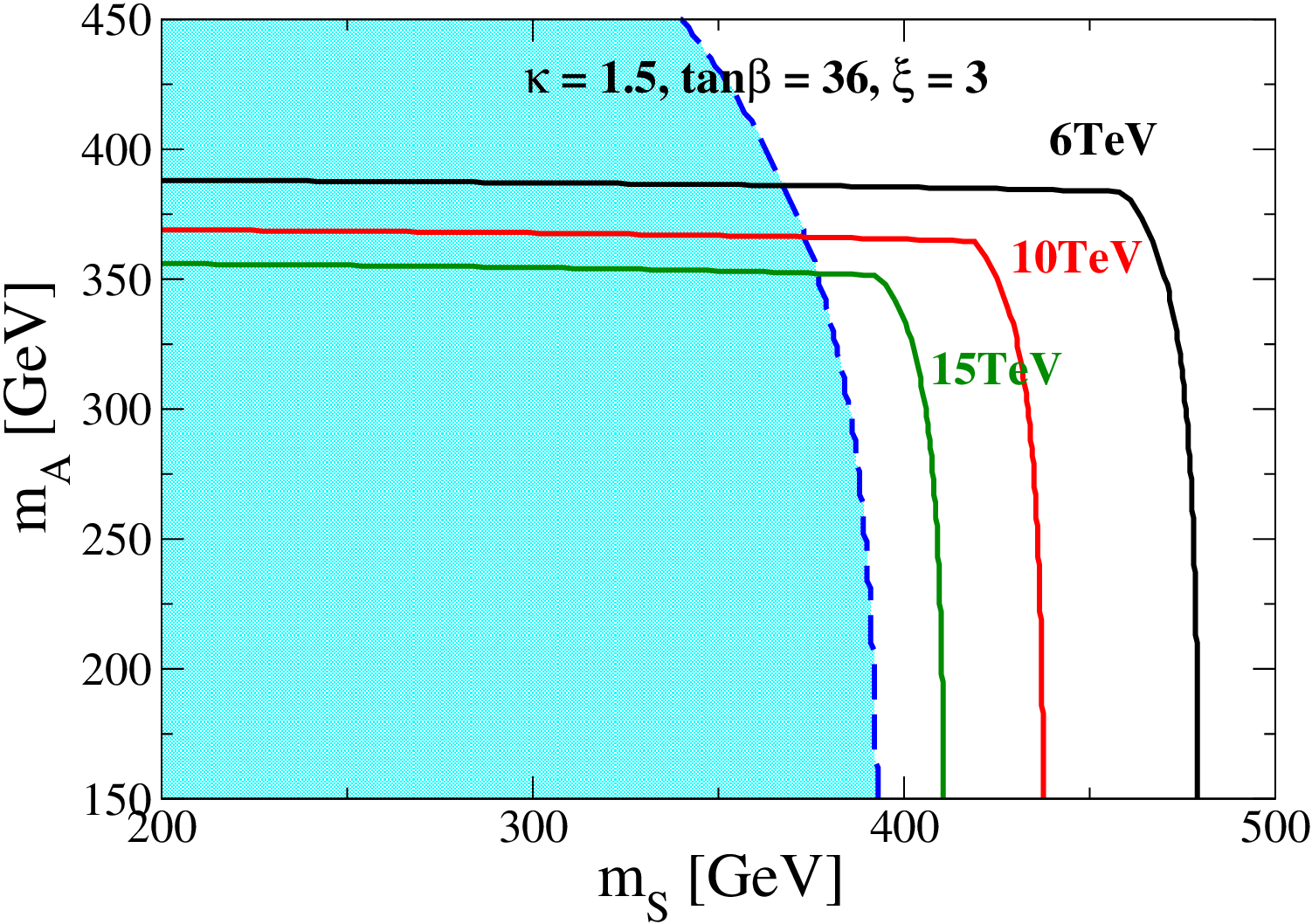}
\caption{Contour plots for the scale where condition of perturbativity is broken are shown 
         in the $m_S$-$m_A$ plane in the case of $(\kappa,\tan\beta)=(1.2, 45)$ (left figure), 
         and $(\kappa,\tan\beta)=(1.5, 36)$  (right figure).   
         The region excluded by the vacuum stability condition is also shown as the shaded area.
         The constant $\xi$ is taken to be 3 at the scale of $m_S$.}
\label{result1}
\vspace{1cm}
\includegraphics[width=75mm]{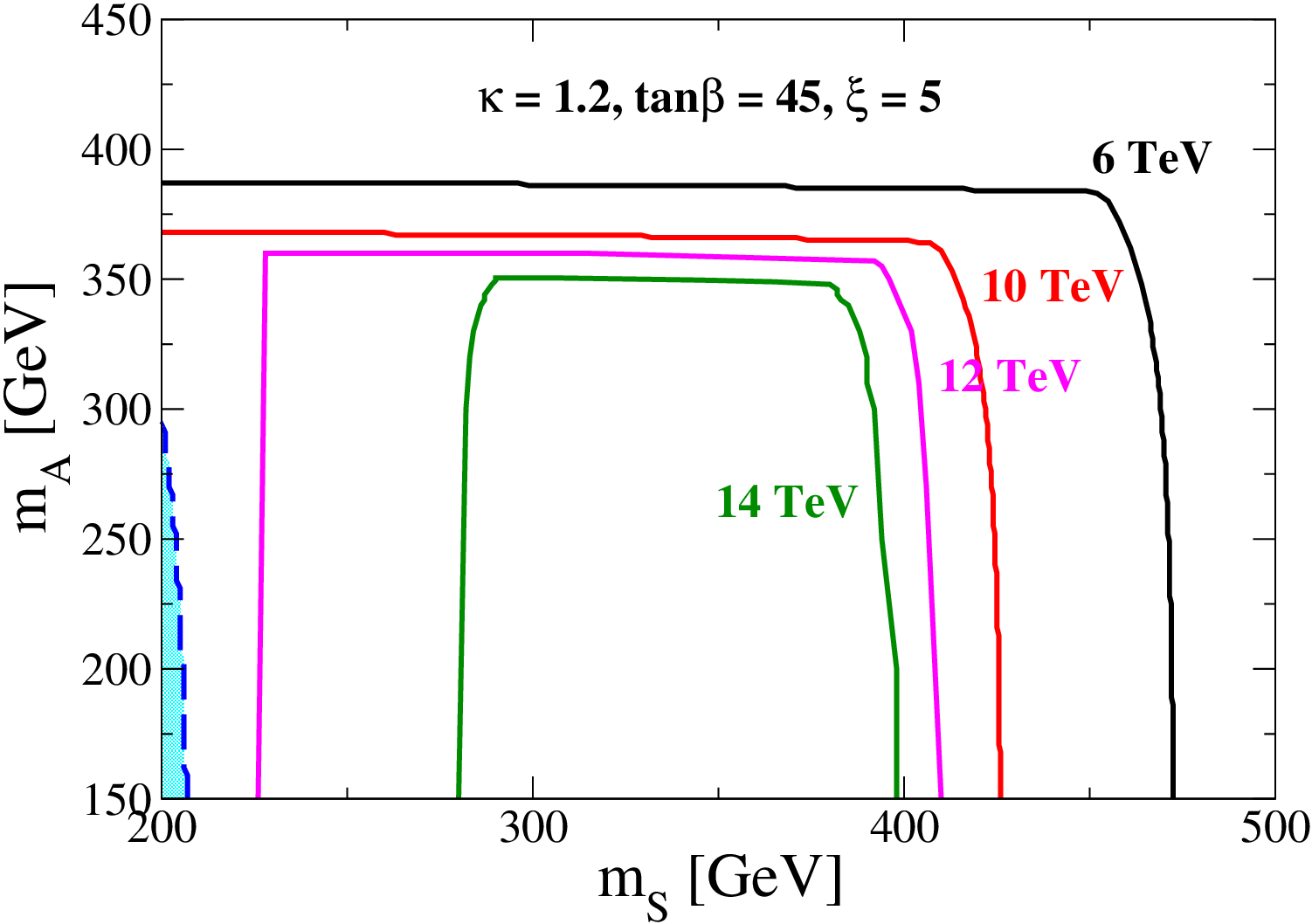}
\includegraphics[width=75mm]{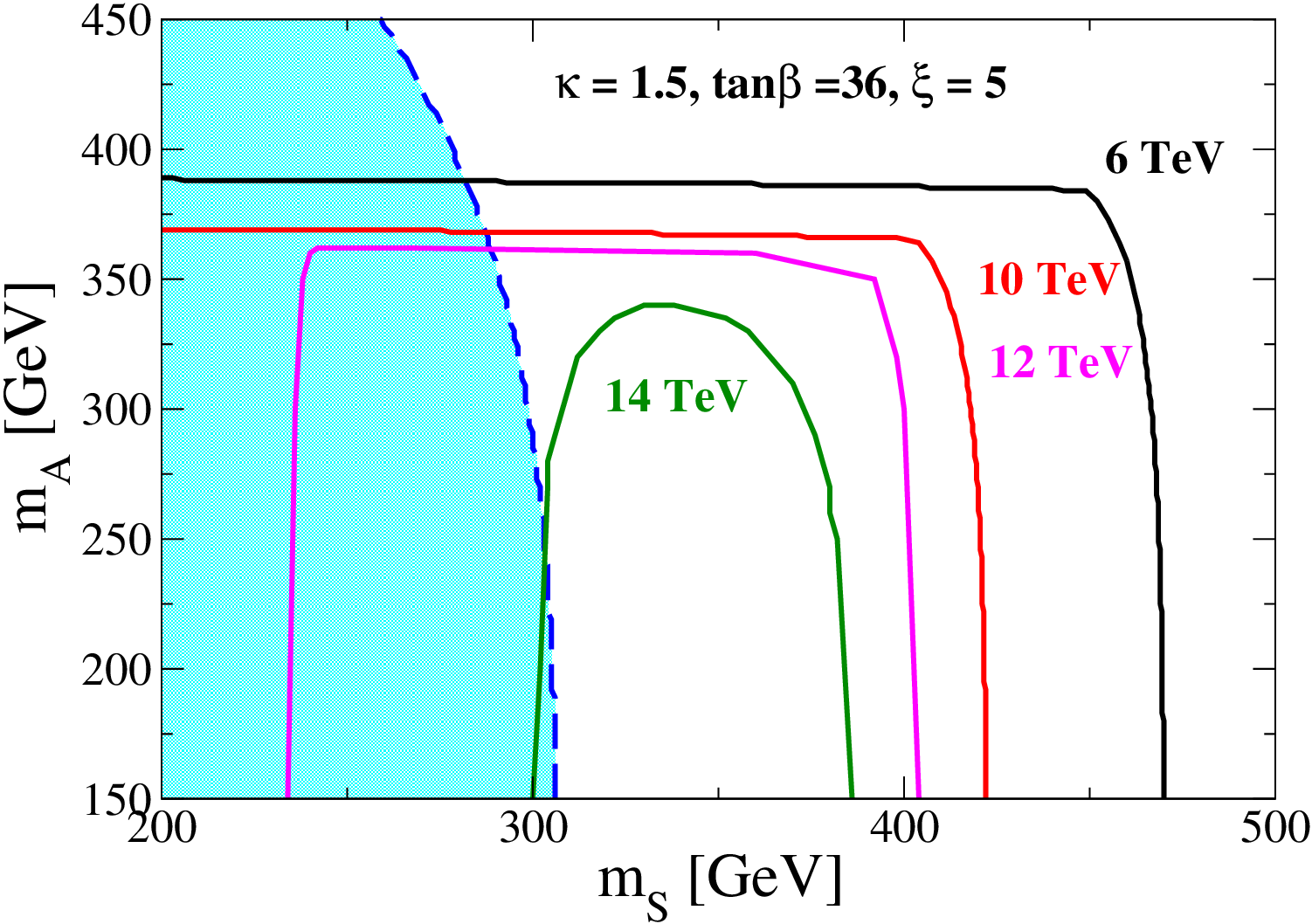}
\caption{Contour plots for the scale where condition 
 of perturbativity is broken are shown 
         in the $m_S$-$m_A$ plane in the case of $(\kappa,\tan\beta)=(1.2, 45)$ (left figure), 
         and $(\kappa,\tan\beta)=(1.5, 36)$  (right figure).   
         The region excluded by the vacuum stability condition is also shown as the shaded area.
         The constant $\xi$ is taken to be 5 at the scale of $m_S$.}
\label{result2}
\end{figure}

The initial values for the scalar coupling constants 
in the Higgs sector are shown in Table.~\ref{constants}.  
We note that the initial value of $\lambda_4$, $\lambda_5$ and $\rho_2$ 
are determined by given values for the masses of $A$ and $S^\pm$ 
using Eqs.~(\ref{mA}), (\ref{mHpm}) and (\ref{mS}).
The rest parameter $\xi$ (the coupling constant for $|S^-|^2\eta^2$) 
is taken as $\xi = 3$ and $5$. 
The results in the case with $\xi=3$ is shown in Fig.~\ref{result1} for 
$\kappa=1.2$ (left figure) and $\kappa=1.5$ (right figure), while those with $\xi=5$ 
is in Fig.~\ref{result2} for the same values of $\kappa$.  

In Fig.~\ref{result1}, the shaded area in the figure is excluded due to the vacuum stability condition 
in Eq.~(\ref{vs_kappa}).  
In this area, the condition is not satisfied already at the electroweak scale, 
so that the excluded region is independent of $\Lambda$. 
The vacuum stability bound become stronger for a larger value of $\kappa$, 
although the area compatible with both theoretical conditions with $\Lambda=10$ TeV 
still exists for $\kappa=1.5$. 
On the other hand,  
the bound from perturbativity depends on $\Lambda$. 
In Fig.~\ref{result1}, the contour plots for $\Lambda=6$, $10$ and $15$ TeV, the scales 
where one of the coupling constants blows up and breaks the 
condition of perturbativity are shown for the case of $\xi=3$ 
in the $m_S$-$m_A$ plane.   
We find that there is the parameter region which satisfies both the conditions of 
vacuum stability and perturbativity with the blow-up scale to be above $\Lambda=10$ TeV.  
The area of the vicinity of $m_S \sim 400$ GeV and $m_A < 350$ GeV can also be 
consistent from the theoretical bounds.
We stress that this parameter region is favored for phenomenologically 
successful scenarios for neutrino masses, relic abundance 
for the dark matter, and the strongly first order phase transition. 

The similar figures but with $\xi=5$ are shown in Fig.~\ref{result2}. 
The contour plots are for $\Lambda=6$, $10$, $12$ and $14$ TeV in the $m_S$-$m_A$ plane.   
We find that there is the parameter region which satisfies both the conditions of 
vacuum stability and perturbativity with the blow-up scale to be above $\Lambda=10$ TeV.  
The vacuum stability bound is more relaxed as compared to that for $\xi=3$, while 
the bound from perturbatibity becomes rather strict.  
In the regions with $m_S < 400$ GeV, the running coupling constants blow up earlier 
than the case with $\xi=3$, because of the threshold effect at the scale $\mu=m_S$, above 
which the running of $\lambda_\eta$ becomes enhanced by the loop contribution of $S^\pm$. 
In the area of $300$ GeV $< m_S < 400$ GeV and $m_A < 350$ GeV, $\Lambda$  can be 
above $10$ TeV.

\begin{table}[t]
\begin{center}
{\renewcommand\arraystretch{1.3}
\begin{tabular}{|c|c|c|c||c|c|c|}\hline
\multicolumn{4}{|c||}{Masses (GeV)} &\multicolumn{3}{c|}{Inv. masses (GeV)} \\\hline\hline
 $m_H^\pm$ &  $m_H$ & $m_h$& $m_\eta$&$M$ &$\mu_S$ &$\mu_\eta$ \\\hline
100 & 100 & 120 & 50 & 100 & 200 & 30 \\\hline
\end{tabular}}
\caption{Values for the scalar boson masses and the invariant mass parameters.}
\label{invariantmass}
\end{center}
\end{table}

\begin{table}[t]
\begin{center}
{\renewcommand\arraystretch{1.3}
\begin{tabular}{|c|c|c|c|c|c|c|c|c||c|}\hline
\multicolumn{9}{|c|}{Scalar couplings} &\multicolumn{1}{||c|}{} \\\hline\hline
$\lambda_1(m_Z)$ &$\lambda_2(m_Z)$ &$\lambda_3(m_Z)$& $\kappa(m_S)\tan\beta$&$\rho_1(m_S)$ &$\lambda_S(m_S)$ &$\sigma_1(m_Z)$ &$\sigma_2(m_Z)$&$\lambda_\eta(m_Z)$&$\sin(\beta-\alpha)$ \\\hline
0.24&0.24&0.24& 54 & 0.1 & 2 & 0.05 &0.05& 3 & 1\\\hline
\end{tabular}}
\caption{Values for the scalar coupling constants and the mixing angles.}
\label{constants}
\end{center}
\end{table}

\section{Conclusion}

We have discussed theoretical constraints on the parameter space under 
the conditions from vacuum stability and triviality 
in the three-loop radiative seesaw model with TeV-scale right-handed 
neutrinos which are odd under the $Z_2$ parity. 
It has been found that the model can be consistent up to the scale 
above 10 TeV in the parameter region which satisfies the neutrino data, 
the LFV data, the thermal relic abundance of dark matter 
as well as the requrement from the strongly first order phase transiton. 
We also reanalyzed the constraint from the LFV data. 
The data from $\mu\to eee$ is found to be more severer than that from 
$\mu\to e\gamma$.   
Our analysis here has been restricted in the case where the CP 
violation is neglected just for simplicity. 
The analysis for consistency with including the CP violating phase, 
which is necessary for finally producing the observed asymmetry of 
baryon number at the electroweak phase transition, will be done in near future.


\vspace{5mm}
\noindent
{\it Acknowledgments}

The work of MA was supported in part by Grant-in-Aid for Young Scientists (B) no.
22740137, that of SK was supported in part by Grant-in-Aid for Scientific Research (A) no.
22244031 and (C) no. 19540277, that of KY was supported by Japan Society for the Promotion of Science (JSPS Fellow (DC2)).

\section*{Appendix}
The full set of the beta functions for RGEs in the model in Ref.~\cite{aks-prl} is given at the one-loop level by
\begin{align}
\beta(g_s)&=\frac{1}{16\pi^2}\left[-7g_s^3\right], \\
\beta(g)&=\frac{1}{16\pi^2}\left[-3g^3\right], \\
\beta(g')&=\frac{1}{16\pi^2}\Big[-\frac{22}{3}g^{'3}\Big],\\
\beta(y_t)&=\frac{1}{16\pi^2}\Big[-8y_tg_s^2-\frac{9}{4}g^2y_t-\frac{17}{12}g^{'2}y_t+\frac{9}{2}y_t^3+\frac{3}{2}y_ty_b^2\Big], \\
\beta(y_b)&=\frac{1}{16\pi^2}\Big[-8y_tg_s^2-\frac{9}{4}g^2y_t-\frac{5}{12}g^{'2}y_t+\frac{9}{2}y_b^3+\frac{3}{2}y_t^2y_b\Big], \\
\beta(y_\tau)&=\frac{1}{16\pi^2}\Big[-\frac{9}{4}g^2y_\tau-\frac{15}{4}g^{'2}y_\tau+\frac{5}{2}y_\tau^3\Big], \\
\beta(h_i^\alpha)&=\frac{1}{16\pi^2}\Big[-5g^{'2}h_i^\alpha+\frac{1}{2}h_i^\alpha\sum_j(h_j^\alpha)^2+\frac{1}{2}h_i^\alpha\sum_\beta(h_i^\beta)^2+h_i^\alpha\sum_{j,\beta}(h_j^\beta)^2\Big], \\
\beta(\lambda_1)&=\frac{1}{16\pi^2}\Big[12\lambda_1^2+4\lambda_3^2+2\lambda_4^2+2\lambda_5^2+4\lambda_3\lambda_4+2\rho_1^2+\sigma_1^2+\frac{9}{4}g^4+\frac{6}{4}g^2g^{'2}+\frac{3}{4}g^{'4}\notag\\
&\hspace{15mm}-4y_\tau^4+(4y_\tau^2-9 g^2-3g^{'2})\lambda_1\Big], \\
\beta(\lambda_2)&=\frac{1}{16\pi^2}\Big[12\lambda_2^2+4\lambda_3^2+2\lambda_4^2+2\lambda_5^2+4\lambda_3\lambda_4+2\rho_2^2+\sigma_2^2+\frac{9}{4}g^4+\frac{6}{4}g^2g^{'2}+\frac{3}{4}g^{'4}\notag\\
&\hspace{15mm}-12y_t^4-12y_b^4+(12y_t^2+12y_b^2-9g^2-3 g^{'2})\lambda_2\Big], \\
\beta(\lambda_3)&=\frac{1}{16\pi^2}\Big[6\lambda_1\lambda_3+2\lambda_1\lambda_4+6\lambda_2\lambda_3+2\lambda_2\lambda_4+4\lambda_3^2+2\lambda_4^2+2\lambda_5^2+2\rho_1\rho_2+\sigma_1\sigma_2+4\kappa^2\notag\\
&\hspace{15mm}+\frac{9}{4}g^4+\frac{3}{4}g^{'4}-\frac{6}{4}g^2g^{'2}+(6y_t^2+6y_b^2+2y_\tau^2-9g^2-3g^{'2})\lambda_3\Big], \\
\beta(\lambda_4)&=\frac{1}{16\pi^2}\Big[2(\lambda_1+\lambda_2+4\lambda_3+2\lambda_4)\lambda_4 
 +8\lambda_5^2-8\kappa^2+3g^2g^{'2} \Big. \nonumber \\
& \hspace*{2cm}\Big. +(6y_t^2+6y_b^2+2y_\tau^2-9g^2-3g^{'2})\lambda_4\Big], \\
\beta(\lambda_5)&=\frac{1}{16\pi^2}\Big[2(\lambda_1+\lambda_2+4\lambda_3+6\lambda_4)\lambda_5+(6y_t^2+6y_b^2+2y_\tau^2-9g^2-3g^{'2})\lambda_5\Big]\\
\beta(\rho_1)&=\frac{1}{16\pi^2}\Big[6\lambda_1\rho_1+4\lambda_3\rho_2+2\lambda_4\rho_2+2\rho_1\lambda_S+4\rho_1^2+\sigma_1\xi+8\kappa^2+3g^{'4}\notag\\
&\hspace{15mm}+(-\frac{15}{2}g^{'2}-\frac{9}{2}g^2+2\sum_{i,\alpha}(h_i^\alpha)^2+2y_\tau^2)\rho_1\Big], \\
\beta(\rho_2)&=\frac{1}{16\pi^2}\Big[6\lambda_2\rho_2+4\lambda_3\rho_1+2\lambda_4\rho_1+2\rho_2\lambda_S+4\rho_2^2+\sigma_2\xi+8\kappa^2+3g^{'4}\notag\\
&\hspace{15mm}+(-\frac{15}{2}g^{'2}-\frac{9}{2}g^2+2\sum_{i,\alpha}(h_i^\alpha)^2+6y_t^2+6y_b^2)\rho_2\Big], 
\end{align}
\begin{align}
\beta(\lambda_S)&=\frac{1}{16\pi^2}\Big[8\rho_1^2+8\rho_2^2+5\lambda_S^2+2\xi^2+24g^{'4}-12g^{'2}\lambda_S \Big. \nonumber \\ 
&\Big. \hspace*{2cm}+4\sum_{i,\alpha}(h_i^\alpha)^2\lambda_S-8\sum_{i,j}\sum_{\alpha,\beta}h_i^\alpha h_i^\beta h_j^\beta h_j^\alpha\Big], \\
\beta(\sigma_1)&=\frac{1}{16\pi^2}\Big[6\lambda_1\sigma_1+(4\lambda_3+2\lambda_4)\sigma_2+\sigma_1\lambda_\eta+2\rho_1\xi+16\kappa^2+(-\frac{9}{2}g^2-\frac{3}{2}g^{'2}+2y_\tau^2)\sigma_1\Big], \\
\beta(\sigma_2)&=\frac{1}{16\pi^2}\Big[6\lambda_2\sigma_2+(4\lambda_3+2\lambda_4)\sigma_1+\sigma_2\lambda_\eta+2\rho_2\xi+16\kappa^2\Big. \nonumber\\ 
& \hspace*{2cm} \Big. +(-\frac{9}{2}g^2-\frac{3}{2}g^{'2}+6y_t^2+6y_b^2)\sigma_2\Big], \\
\beta(\lambda_\eta)&=\frac{1}{16\pi^2}\Big[12(\sigma_1^2+\sigma_2^2)+3\lambda_\eta^2+6\xi^2\Big], \\
\beta(\kappa)&=\frac{1}{16\pi^2}\kappa\Big[2\lambda_3-2\lambda_4+2\xi+2\sigma_1+2\sigma_2+2\rho_1+2\rho_2+\sum_{\alpha,i}(h_i^\alpha)^2\Big. \nonumber \\
& \hspace*{2cm} \Big.-\frac{9}{2}g^2-\frac{9}{2}g^{'2}+3y_t^2+3y_b^2+y_\tau^2\Big], \\
\beta(\xi)&=\frac{1}{16\pi^2}\Big[4\rho_1\sigma_1+4\rho_2\sigma_2+2\lambda_S\xi+\lambda_\eta\xi+4\xi^2-6g^{'2}\xi+2\sum_{\alpha,i}(h_i^\alpha)^2\xi\Big]. 
\end{align}

\end{document}